\documentclass[12pt]{article}
\usepackage{graphicx}
\usepackage {hyperref}
\usepackage {color}
\usepackage{amsmath}
\usepackage{amsfonts}
\usepackage{amssymb}
\usepackage{epstopdf}
\usepackage{color}
\usepackage{amsmath,amssymb}
\usepackage{slashed}
\usepackage{xcolor} 
\usepackage{url}
\usepackage{multirow}
\usepackage[noadjust]{cite}
\usepackage{filecontents}
\usepackage{lineno}
\usepackage{subfig}
\usepackage[italic]{hepnames}
\usepackage{perpage} 
\MakePerPage{footnote} 

\newcommand\pubnumber{SNSN-323-63}
\newcommand\pubdate{\today}

\def\institute{Interuniversity Institute for High Energies (IIHE),\\
Physique des particules \'el\'ementaires,\\
Universit\'e Libre de Bruxelles, ULB, 1050, Brussels, Belgium}

\def\Title#1{\begin{center} {\Large #1 } \end{center}}
\def\Author#1{\begin{center}{ \sc #1} \end{center}}
\def\Address#1{\begin{center}{ \it #1} \end{center}}

\newcommand\pubblock{\rightline{\begin{tabular}{l} \pubnumber\\
         \pubdate  \end{tabular}}}
\newenvironment{Abstract}{\begin{quotation}  }{\end{quotation}}
\newenvironment{Presented}{\begin{quotation} \begin{center} 
             PRESENTED AT\end{center}\bigskip 
      \begin{center}\begin{large}}{\end{large}\end{center} \end{quotation}}

\def\beq{\begin{equation}}
\def\eeq#1{\label{#1}\end{equation}}
\def\eeqn{\end{equation}}

\def\beqa{\begin{eqnarray}}
\def\eeqa#1{\label{#1}\end{eqnarray}}
\def\eeqan{\end{eqnarray}}

\let\bar=\overbar

\def\Dslash{\not{\hbox{\kern-4pt $D$}}}
\def\dslash{\not{\hbox{\kern-2pt $\del$}}}

\def\msb{{\bar{\ssstyle M \kern -1pt S}}}

\begin{document}
\begin{titlepage}
\pubblock

\vfill
\Title{Search for anomalous tq$\gamma$ FCNC Couplings in direct single top quark production at the LHC}
\vfill
\Author{ Reza Goldouzian and Barbara Clerbaux}
\Address{\institute}
\vfill
\begin{Abstract}
In this paper we propose and study direct single top quark production channel for probing the tq$\gamma$ flavor-changing neutral current (FCNC) interactions at the LHC. We reinterpret the experimental results of a search for direct single top quark production performed by ATLAS collaboration at 8 TeV to set upper bounds on the FCNC top quark decay branching fractions through the proposed channel, resulting in ${\cal B}$(t$\rightarrow$ u$\gamma$) $<$ 0.05\% and ${\cal B}$(t$\rightarrow$ c$\gamma$) $<$ 0.14\%. Finally, the expected sensitivity at 13 TeV for various integrated luminosity scenarios is estimated.
   
\end{Abstract}
\vfill
\begin{Presented}
$9^{th}$ International Workshop on Top Quark Physics\\
Olomouc, Czech Republic,  September 19--23, 2016
\end{Presented}
\vfill
\end{titlepage}
\def\thefootnote{\fnsymbol{footnote}}
\setcounter{footnote}{0}

\section{Introduction}

In the standard model (SM),  top quark flavour-changing neutral current (FCNC) interactions are not present at tree level and are suppressed at higher order quantum loop levels. In some scenarios beyond the SM, an enhancement of the top quark FCNC processes is predicted \cite{AguilarSaavedra:2004wm}. Therefore, top quark FCNC processes are known as possible windows to explore new physics.

In order to evaluate the contribution of the FCNC processes independent of the new physics models, an effective Lagrangian approach is employed. The most general effective Lagrangian of dimension six operators which describes the tq$\gamma$ FCNC interaction can be written as\footnote{The parameter description can be found in ref.\cite{Goldouzian:2016mrt}.} \cite{AguilarSaavedra:2004wm}:
\begin{eqnarray}
-\mathcal{L}_{eff} = e \frac{\kappa_{\text{q}\gamma}}{\Lambda} \bar{q} i\sigma^{\mu\nu}k_{\nu} [\gamma_{L} P_{L}+\gamma_{R} P_{R}]tA_{\mu} +  h.c.
\label{lagrangy}
\end{eqnarray}

Various signal channels are proposed and investigated to probe  the tq$\gamma$ FCNC couplings in proton-proton collisions \cite{AguilarSaavedra:2004wm}.
The LHC is known not only for proton-proton collisions but also for photon-proton and photon-photon collisions.  
Therefore, in the presence of tq$\gamma$ FCNC couplings, the radiated photon from one proton can interact with the up or charm quark from the other proton and can produce a single top quark without any additional object (direct single top quark production).

In previous related studies \cite{deFavereaudeJeneret:2008hf,Sun:2014qoa,Koksal:2013fta}, direct single top quark production channel through elastic photon interaction was proposed for probing tq$\gamma$ FCNC couplings. The requirement of elastic photon emission provides a clean signature for signal events and reduces SM backgrounds since the proton can be detected in the very forward spectrometers close to the beam line after photon radiation. On the other hand, tagging deflected proton would be inefficient when the number of pileup interactions increases.  In this paper, we propose and study direct single top quark production through both elastic and inelastic photon emissions.
 
\section{Signal cross section}
Direct single top quark production due to tq$\gamma$ FCNC interactions can originate from elastic and inelastic photon emissions. In order to describe the signal cross section, we separate these two production mechanisms. Equivalent photon approximation (EPA) approach is used for the elastic photon emission part in which the cross section of $\gamma$p $\rightarrow$ t is folded with the photon flux radiated from protons \cite{Budnev:1974de}. Photon distribution functions  in the proton are used to calculate the signal cross section for the inelastic photon emission part\footnote{MRST\_QED, CT14\_QED and NNPDF23\_QED are used.}.

The effective Lagrangian is implemented into the {\sc FeynRules} program and is used in {\sc MadGraph\_}a{\sc mc@NLO} to calculate cross sections and to generate events.  In table \ref{tab:tCS}, cross sections of direct single top quark production through elastic and inelastic photon radiation as a function of tq$\gamma$  anomalous couplings $\kappa_{\text{u}\gamma}$ and $\kappa_{\text{c}\gamma}$ for 8 TeV and 13 TeV are summarized.

\begin{table}[h]
\centering
\begin{tabular}{l|ccc}
Photon emission  & PDF set    & $\sqrt{s}$ & Cross section (pb)     \\
\hline
Elastic  & EPA    & 8 TeV  & 225 $\kappa_{\text{u}\gamma}^2$ + 68  $\kappa_{\text{c}\gamma}^2$    \\
Elastic  & EPA    & 13 TeV  & 366 $\kappa_{\text{u}\gamma}^2$ + 140  $\kappa_{\text{c}\gamma}^2$     \\
Inelastic  & MRST\_QED    & 8 TeV  &  829 $\kappa_{\text{u}\gamma}^2$ +  267 $\kappa_{\text{c}\gamma}^2$     \\
Inelastic  & CTEQ\_QED    & 8 TeV  &  505 $\kappa_{\text{u}\gamma}^2$ +   133 $\kappa_{\text{c}\gamma}^2$     \\
Inelastic  & NNPDF\_QED    & 8 TeV  & 687 $\kappa_{\text{u}\gamma}^2$ + 294  $\kappa_{\text{c}\gamma}^2$     \\
Inelastic  & MRST\_QED  & 13 TeV  &  1392 $\kappa_{\text{u}\gamma}^2$ +  546  $\kappa_{\text{c}\gamma}^2$     \\
Inelastic  & CTEQ\_QED   & 13 TeV  & 905 $\kappa_{\text{u}\gamma}^2$ +  295 $\kappa_{\text{c}\gamma}^2$     \\
Inelastic  & NNPDF\_QED    & 13 TeV  &  1082 $\kappa_{\text{u}\gamma}^2$ + 546  $\kappa_{\text{c}\gamma}^2$     \\
\hline
\end{tabular}
\caption{Total cross sections at center of mass energies of 8 and  13 TeV of the pp $\rightarrow$ t for both elastic and inelastic photon emissions using various PDF sets as a function of the tq$\gamma$ FCNC couplings.}
\label{tab:tCS}
\end{table}

In addition to the pp $\rightarrow$ t channel, tq$\gamma$ FCNC couplings leads to other final states including single top quark. These processes are pp $\rightarrow$ t+jet, pp $\rightarrow$ t$\gamma$ and pp $\rightarrow$ t$\bar{\text{t}}$ $\rightarrow$  t$\gamma$q in which top quark is accompanied with extra objects. The corresponding cross sections are shown in table \ref{tjetCS}. Exclusive analyses could be used for probing the tq$\gamma$ FCNC couplings in the mentioned channels. However, we will include their contributions to the direct single top quark final state to increase the sensitivity.

\begin{table}[h]
\centering
\begin{tabular}{l|cc}
    & $\sqrt{s}$ = 8 TeV & $\sqrt{s}$ = 13 TeV      \\
\hline
pp $\rightarrow$ t+jet    & 519 $\kappa_{\text{u}\gamma}^2$ +  227 $\kappa_{\text{c}\gamma}^2$   & 998 $\kappa_{\text{u}\gamma}^2$ + 501  $\kappa_{\text{c}\gamma}^2$  \\
pp $\rightarrow$ t$\gamma$    &  96 $\kappa_{\text{u}\gamma}^2$ +  11 $\kappa_{\text{c}\gamma}^2$  & 235 $\kappa_{\text{u}\gamma}^2$ + 36  $\kappa_{\text{c}\gamma}^2$    \\
pp $\rightarrow$ t$\bar{\text{t}}$ $\rightarrow$  t$\gamma$q &  112.7 $\kappa_{\text{q}\gamma}^2$  &  370.76 $\kappa_{\text{q}\gamma}^2$  \\
\hline
\end{tabular}
\caption{The total cross section at center of mass energies of 8 and  13 TeV for the pp $\rightarrow$ t+jet ($p_T^{\text{jet}}>10$ GeV), pp $\rightarrow$ t$\gamma$ ($p_T^{\gamma}>10$ GeV) and pp $\rightarrow$ t$\bar{\text{t}}$ $\rightarrow$  t$\gamma$q as a function of the tq$\gamma$ FCNC couplings.}
\label{tjetCS}
\end{table}


\section{Limits}

To estimate the sensitivity of the proposed signal channel for constraining the tq$\gamma$ FCNC couplings, we reinterpret the results of an experimental search with the same final state performed by the ATLAS collaboration using 20.3 fb$^{-1}$ of  data collected at 8 TeV \cite{Aad:2015gea}. In ref.\cite{Aad:2015gea}, the direct single top quark production process is used to probe top-quark-gluon FCNC interactions. No evidence of signal events is observed and an upper limit is set on the signal cross section to 3.4 pb.   

\begin{figure}[t]
\centering 
\includegraphics[width=.48\textwidth]{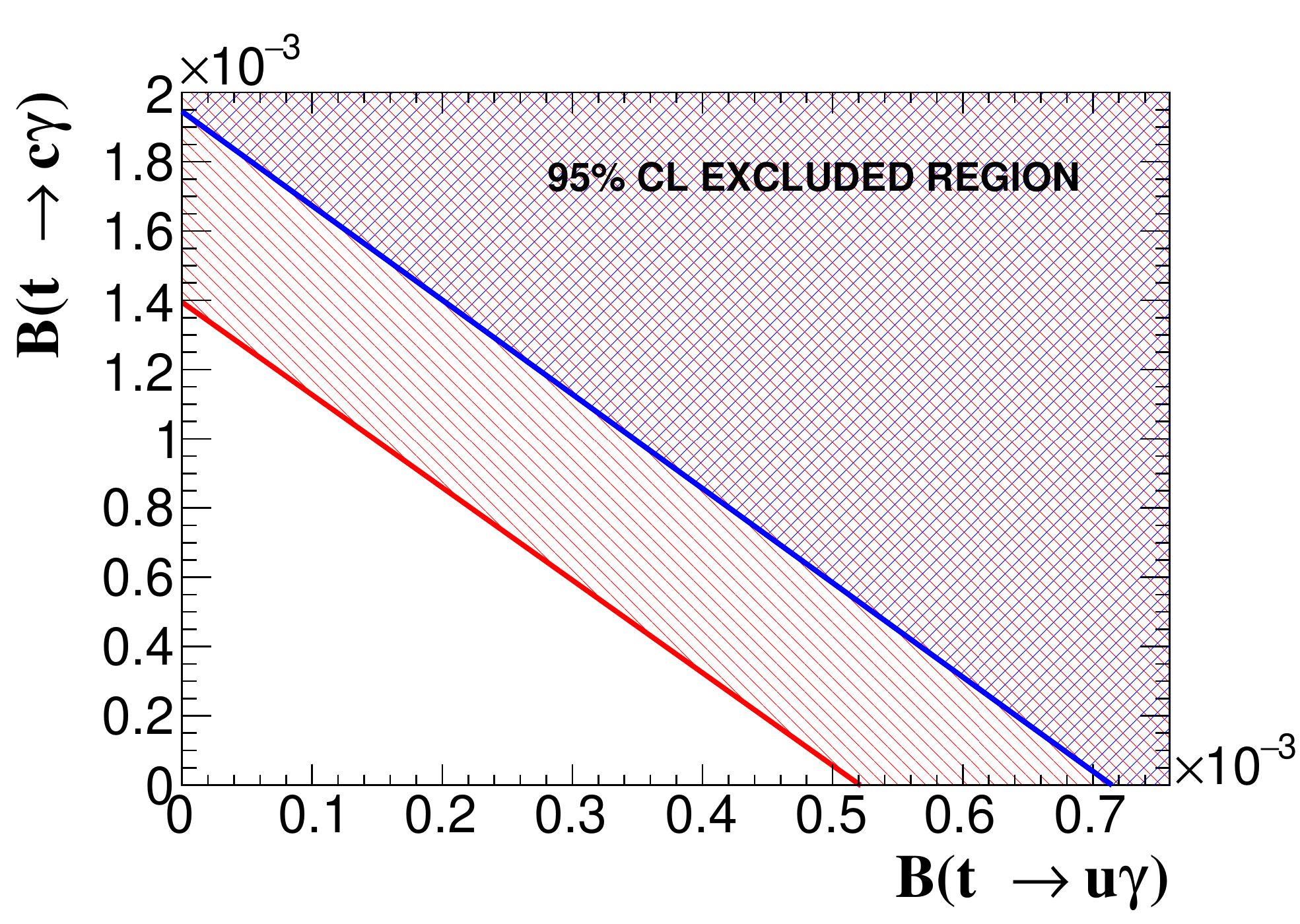}
\includegraphics[width=.48\textwidth]{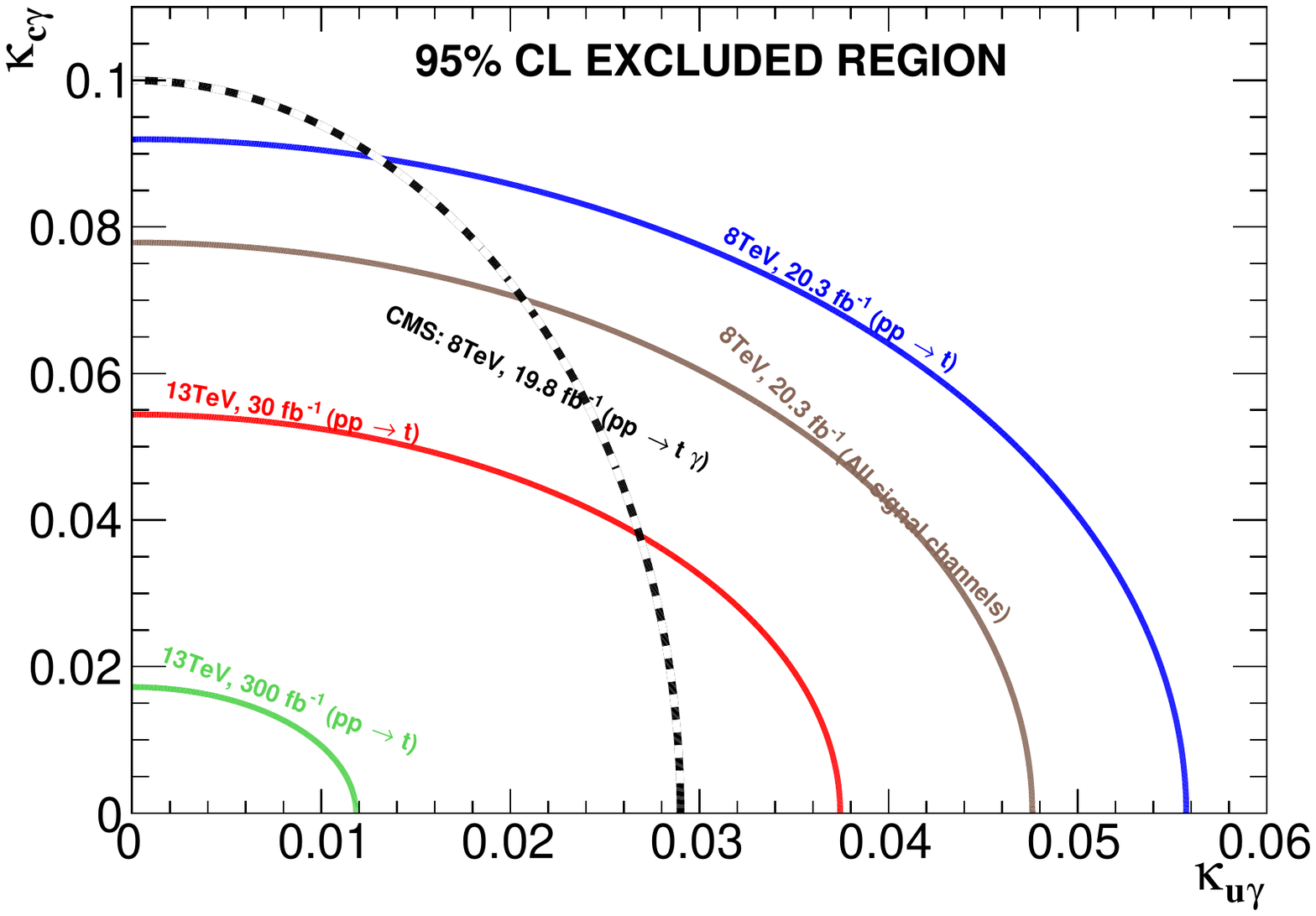}
\hfill
\caption{Left: Excluded region at 95\% CL on the branching fraction plane of (${\cal B}$(t $\rightarrow$ c$\gamma)$, ${\cal B}$(t $\rightarrow$ u$\gamma)$) using pp $\rightarrow$ t  process (blue hatched regions) and all signal channels (red hatched regions). Right: Excluded region at 95\% CL on the coupling constant plane of ($\kappa_{\text{c}\gamma}$, $\kappa_{\text{u}\gamma}$) obtained via pp $\rightarrow$ t channel and all signal channels at 8 TeV (blue and brown curves). The most stringent result obtained by the CMS collaboration via pp $\rightarrow$ t$\gamma$ at 8 TeV is shown with dashed black curve \cite{Khachatryan:2015att}. The projection results to 13 TeV are shown with red and green curves.  }
\label{limitplots}
\end{figure}

In the present analysis signal events are generated with {\sc MadGraph\_}a{\sc mc@NLO} and are passed into the {\sc Pythia} 8 for hadronisation and showering. Detector effects are simulated with {\sc Delphes} program. The ATLAS analysis is closely followed for setting selection cuts\footnote{The full selection cuts can be found in ref.\cite{Goldouzian:2016mrt}.}. The fraction of survived events after the selection cuts is calculated to be 2.4\% and 2\%  for tq$\gamma$ and tqg signal samples, respectively. 

The upper limit reported by ATLAS collaboration is extracted from the analysis of a multivariate output. The distributions of multivariate input variables are checked for tq$\gamma$ and tqg signal samples and similar behavior is observed.  Constraints obtained on the tq$\gamma$ FCNC couplings are shown in table \ref{FCNCresults} and figure \ref{limitplots} (left). 

\begin{table}[h]
\centering
\begin{tabular}{l|cccc}
 & $\kappa_{\text{u}\gamma}$ & ${\cal B}$(t $\rightarrow$ u$\gamma)$    & $\kappa_{\text{c}\gamma}$ & ${\cal B}$(t $\rightarrow$ c$\gamma)$      \\
\hline
pp $\rightarrow$ t channel & 0.056 & 0.07\%    &  0.092 &     0.19\%   \\
All signal channels & 0.048\    &  0.05\% & 0.078  &  0.14\%  \\
\hline
\end{tabular}
\caption{Upper limits at 95\% confidence level on the tq$\gamma$ FCNC couplings and the branching fraction of top quark FCNC decays through pp $\rightarrow$ t only process and the combination of pp $\rightarrow$ t, pp $\rightarrow$ t+jet, pp $\rightarrow$ t$\gamma$ and pp $\rightarrow$ t$\bar{\text{t}}$ $\rightarrow$  t$\gamma$q processes. }
\label{FCNCresults}
\end{table}

At 13 TeV, the LHC will provide more data and the same analysis can be done by the CMS and ATLAS collaborations. To estimate the reach of the mentioned analysis at 13 TeV, we extrapolate the 8 TeV results. In the projection procedure, it is assumed that the experimental searches can reach the same upper bound on the signal cross section as the one at 8 TeV bound. Considering the increase of signal cross section from 8 TeV to 13 TeV, upper limits on the tq$\gamma$ FCNC couplings are estimated for scenarios of total integrated luminosities of 30 fb$^{-1}$ and 300 fb$^{-1}$. The obtained results at 13 TeV are shown in figure \ref{limitplots} (right).

\section{Conclusions}

Direct single top quark production process is proposed and studied as a sensitive channel for probing tq$\gamma$ FCNC couplings in proton-proton collisions. Experimental results at 8 TeV are used to set first experimental constraints on the tq$\gamma$ FCNC couplings through the direct single top quark production channel. The constraints obtained on the tq$\gamma$ FCNC couplings are comparable to the most stringent limits. It was shown that this channel is expected to be more sensitive at 13 TeV especially for probing tc$\gamma$ FCNC coupling.

\end{document}